\documentclass[pra,aps,twocolumn,epsfig,superscriptaddress,showpacs]{revtex4}
\usepackage{bm}
\usepackage{amsfonts}
\usepackage[dvips]{graphicx}
\usepackage{mathrsfs}
\usepackage[intlimits]{amsmath}
\usepackage[colorlinks=false,dvipdfm]{hyperref}

\begin{document}

\title{Detect genuine multipartite entanglement in the one-dimensional transverse-field
Ising model}
\author{Dong-Ling Deng}
\affiliation{Theoretical Physics Division, Chern Institute of Mathematics, Nankai
University, Tianjin 300071, People's Republic of China}
\author{Shi-Jian Gu}
\affiliation{Department of Physics and ITP, The Chinese University
of Hong Kong, Hong Kong, China}

\author{Jing-Ling Chen}
\email{chenjl@nankai.edu.cn}
\affiliation{Theoretical Physics Division, Chern Institute of Mathematics, Nankai
University, Tianjin 300071, People's Republic of China}

\date{\today}

\begin{abstract}
Recently Seevinck and Uffink argued that genuine multipartite
entanglement (GME) had not been established in the experiments
designed to confirm GME. In this paper, we use the Bell-type
inequalities introduced by Seevinck and Svetlichny [Phys. Rev. Lett.
\textbf{89}, 060401 (2002)] to investigate the GME problem in the
one-dimensional transverse-field Ising model. We show explicitly
that the ground states of this model violate the inequality when the
external transverse magnetic field is weak, which indicate that the
ground states in this model with weak magnetic field are fully
entangled. Since this model can be simulated with nuclear magnetic
resonance, our results provide a fresh approach to experimental test
of GME.
\end{abstract}

\maketitle

Quantum entanglement is central to the foundations of quantum mechanics and
has been playing a vital role in quantum communication, information and
computation processing, such as quantum teleportation \cite%
{QuantumTele-Exp,1993Bennett-G.Brassard}, quantum cryptographic schemes~\cite%
{1991Ekert}, quantum parallelism \cite{1997Deutsch}, dense coding~\cite%
{1996Mattle,1992Bennett-S.J.Wiesner}, decoherence in quantum computers, and
the evaluation of quantum cryptographic schemes~\cite{1997Fuchs}. While
two-particle entanglement is well-known, multipartite entanglement is far
from comprehensively understanding not only because the classification of
different types of this form of entanglement is still an open problem~\cite%
{1995Popescu-Lewenstein}, but also it requires different conditions for
actual experimental confirmation. Extensive efforts related to this issue
have been made both theoretically and experimentally~\cite%
{multi-qubits-By-Quadratic-BI,2002Seevinck,
2002Collins,multipartite-entanglement-experiment}. In fact, recent
experiments using photons and atom-cavity techniques~\cite%
{multipartite-entanglement-experiment} have claimed experimental conformation
of genuine multipartite entanglement (GME). Nevertheless, Seevinck and Uffink
argued that GME had not been established in these experiments by pointing out a
possible loophole problem \cite{2001Seevinck}. Here, we use the Bell-type
inequalities introduced by Seevinck and Svetlichny~\cite{2002Seevinck} to
investigate the GME problem in the one-dimensional transverse-field Ising
model. This model can be simulated experimentally, for instance with nuclear
magnetic resonance (NMR)\cite{JFZhang2008}, thus leading to a fresh approach to
experimental test of GME.

%
%
%

The term \emph{genuine multipartite entanglement} refers to states in which
none of the parties can be separated from any other party in a mixture of
product states. Generally speaking, it is difficult to determine whether a
quantum state bears GME or is just partially entangled since, in principle,
we should write down all the possible decompositions of the state.
Fortunately, fruitful results concerning this problem have provided
sufficient conditions for detecting GME. In $1987$, Svetlichny derived, for
the first time, a Bell-type inequality to distinguish three-body from
two-body entanglement~\cite{1987Svetlichny}. Then, in $2002$, this result
was generalized to multi-qubit systems~\cite{2002Seevinck,2002Collins}.


An experimental feasible model that might possess GME is the one-dimensional
transverse-field Ising model~\cite{IsingM}, which has a simple and clear
physical picture and can be solved exactly~\cite{IsingM-ExactS}. It is one of
the most important models in the field of condensed matter physics and is often
used as a starting model to test new physical ideas and approaches. In this
paper, we investigate the GME problem in this model based on the
Bell-type inequalities introduced by Seevinck and Svetlichny~\cite%
{2002Seevinck}. We restrict our study on the ground states of this model and
find that these states violate the inequalities when the external transverse
magnetic field is weak. We also find that the violations might decrease as
the external magnetic field gets stronger and stronger. The more particles
involved in the model, the faster the violations decrease.


To begin with, let first briefly review the the $N$-particle Bell-type
inequalities presented by Seevinck and Svetlichny~\cite{2002Seevinck}, which
are useful for detecting GME. Consider an experimental situation involving $%
N $ particles in which two measurements $X^{[j]}_1$ and $X^{[j]}_2$ ($%
j=1,\cdots,N$) can be performed on each particle. Each of the measurement
has two possible outcomes: $X^{[j]}_{k_j}=\pm1$ ($k_j=1,2$). In a specific
run of the experiment the correlations between all $N$ observations can be
represented by the product $\Pi^N_{j=1}X^{[j]}_{k_j}$, then the correlation
function is the average over many runs of the experiment:
\begin{eqnarray}  \label{CorrF}
Q_{k_1\cdots k_N}=\left\langle\Pi^N_{j=1}X^{[j]}_{k_j}\right\rangle
\end{eqnarray}
Based on the partial separability~\cite{2002Seevinck,1987Svetlichny}, or
more generally speaking, on the so called \emph{hybrid local-nonlocal hidden
variables} (HLNHV) models~\cite%
{2004Mitchell-three-particle-nonlocality,2002Collins}, Seevinck and
Svetlichny introduced the following Bell-type inequalities:
\begin{eqnarray}  \label{general-QBI}
I^{[N]}=\frac{1}{2^{N-1}}\sum_{k_1,\cdots,k_N}\mathcal{V}^{\pm}_{%
\kappa(K)}Q_{k_1\cdots k_N}\leq1,
\end{eqnarray}
where $K=(k_1,k_2,\cdots,k_N)$ and $\kappa(K)$ is the number of times of
index $2$ appears in $K$; $\mathcal{V}^{\pm}_{\kappa(K)}$ be a sequence of
signs given by $\mathcal{V}^{\pm}_{\kappa(K)}=(-1)^{\kappa(\kappa\pm1)/2}$.
As analyzed in Ref. \cite{2002Seevinck}, the two inequalities of Eq. (\ref%
{general-QBI}) are equivalent for even $N$ since they are interchanged by a
global change of labels $1$ and $2$. While for odd $N$ this is not the case
and should be considered a priori independent. Moreover, the inequalities of
Eq.~(\ref{general-QBI}) are invariant under a permutation of the $N$
particles. Although the inequalities of Eq.~(\ref{general-QBI}) are derived
from the HLNHV model, they also hold for quantum states that are partially
entangled. Thus, the violations of these inequalities are sufficient
conditions for GME. Actually, in quantum mechanics, the observables could be
spin projections onto unit vectors $X^{[j]}_{k_j}=\mathbf{n}^{[j]}_{k_j}\cdot%
\vec{\sigma}_j$. Here $\mathbf{n}^{[j]}_{k_j}$ is an unit vector and $\vec{%
\sigma}_j=(\sigma^x_j, \sigma^y_j, \sigma^z_j)$ is a vector of local Pauli
operators for the $j$th observer. For any quantum state $\rho$, the quantum
expression of the correlation function reads: $Q_{k_1\cdots k_N}=\mathtt{Tr}%
[\rho X^{[1]}_{k_1}\otimes\cdots\otimes X^{[N]}_{k_N}]$. Then it is
obviously that if the quantum state $\rho$ is partially entangled, the
quantum expression of the correlation function might be factorizable.
Consequently, from the construction of the inequalities~(\ref{general-QBI}),
any partially entangled quantum state might obey these inequalities. For
detailed analysis, please see Ref. \cite{2002Seevinck}. %
%
%
%
%
%
For simplicity and convenience, in this paper, we restrict ourselves to the
case of $\mathcal{V}^-_{\kappa({K})}$ and the case of $\mathcal{V}^+_{\kappa(%
{K})}$ can be analyzed similarly. For the case of $\mathcal{V}^-_{\kappa({K}%
)}$, the inequality for three-qubit system reads:
\begin{eqnarray}  \label{ThreeQubits-BTI}
I^{[3]}&=&\frac{1}{4}(Q_{111}+Q_{112}+Q_{121}+Q_{211}-Q_{122}   \\
&-&Q_{212}-Q_{221}-Q_{222})\leq1.
\end{eqnarray}
Inequality~(\ref{ThreeQubits-BTI}) was first introduced by Svetlichny in $%
1987$ to distinguish three-particle from two-particle entanglement~\cite%
{1987Svetlichny}. Here we apply it in the one-dimensional transverse-field
Ising model.

The Hamiltonian of the one-dimensional transverse-field Ising model with
periodic boundary conditions reads
\begin{eqnarray}  \label{NqubitHam}
H_N=-\sum_{i=1}^{N}(\sigma_i^x\sigma_{i+1}^x+h\sigma_i^z),\quad%
\sigma^x_{N+1}=\sigma^x_1,
\end{eqnarray}
where $h$ is the transverse field and $N$ is the number of spins involved in
the model. As inferred from the model's name, the Hamiltonian describes a
chain of spins with the nearest neighboring Ising interaction along $x$%
-direction, and all spins are subject to a transverse magnetic field $h$
along the $z$-direction. If $h\rightarrow\infty$, the Ising interaction is
neglectable, and all spins are fully polarized along $z$-direction. It is
easy to prove that the ground state for a finite system in the whole region $%
h>0$ is nondegenerate. An illustration of this model is shown in Fig. \ref%
{IsingFig}.
\begin{figure}[tbp]
\includegraphics[width=75mm]{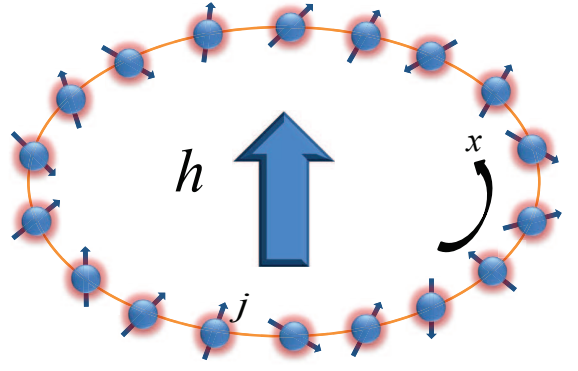}\newline
\caption{(Color online) A sketch of the one-dimensional
transverse-field Ising model with periodic boundary conditions. Here
only eighteen spins are considered. In this model, all spins are
subject to an external field $h$ along $z$ direction and two
arbitrary neighboring spins interact with each other by the Ising
interaction $\protect\sigma^x_j\protect\sigma^x_{j+1}$. }
\label{IsingFig}
\end{figure}

%

To detect GME in this model, let first focus on the three qubit-case, namely
$N=3$. In this case, the Hamiltonian of this model in the matrix form reads:
\begin{eqnarray}
H_3=\left(%
\begin{matrix}
-3h & 0 & 0 & -1 & 0 & -1 & -1 & 0 \\
0 & -h & -1 & 0 & -1 & 0 & 0 & -1 \\
0 & -1 & -h & 0 & -1 & 0 & 0 & -1 \\
-1 & 0 & 0 & h & 0 & -1 & -1 & 0 \\
0 & -1 & -1 & 0 & -h & 0 & 0 & -1 \\
-1 & 0 & 0 & -1 & 0 & h & -1 & 0 \\
-1 & 0 & 0 & -1 & 0 & -1 & h & 0 \\
0 & -1 & -1 & 0 & -1 & 0 & 0 & 3h%
\end{matrix}%
\right).
\end{eqnarray}
Solve the Hamiltonian $H_3$ directly, one can easily find that the ground
state of this Hamiltonian is:
\begin{eqnarray}  \label{3SpinGS}
|\psi_3\rangle_{g}=(\gamma_1|000\rangle+|011\rangle+|101\rangle+|110\rangle)/%
\sqrt{\mathcal{N}_1},
\end{eqnarray}
where $\gamma_1=-1+2h+2\sqrt{1-h+h^2}$ and $\mathcal{N}_1=3+\gamma_1^2$ is
the normalization constant. Our numerical results show that, for small $h$,
the ground state $|\psi_3\rangle_{g}$ violates the Bell-type inequality~(\ref%
{ThreeQubits-BTI}), indicating that the state is a genuine three-particle
entangled state. For instance, denote the unit vector on to which the spin
projected by $\mathbf{n}^{[j]}_{k_j}=(\sin\theta^{[j]}_{k_j}\cos%
\phi^{[j]}_{k_j},
\sin\theta^{[j]}_{k_j}\sin\phi^{[j]}_{k_j},\cos\theta^{[j]}_{k_j})$. The
maximal violation of the inequality~(\ref{ThreeQubits-BTI}) occurs at $h=0$
and the violation is $\sqrt{2}$. To obtain the maximal violation, we can set
$\theta^{[1]}_1=\theta^{[2]}_1=\phi^{[2]}_1=0$, $\theta^{[1]}_2=%
\theta^{[2]}_2=\phi^{[1]}_2=\phi^{[2]}_2= \phi^{[3]}_1=\phi^{[3]}_2=\pi/2$, $%
\phi^{[1]}_1=\pi/3$, $\theta^{[3]}_1=-\pi/4$, and
$\theta^{[3]}_2=\pi/4$. This fact is not surprising because the
ground state $|\psi_3\rangle_g$ with $h=0$ is
equivalent to the maximally entangled state in the GHZ form $%
|\Psi_3\rangle_{max}=\frac{1}{\sqrt{2}}(|000\rangle+|111\rangle)$ due to
local unitary transformations~\cite{2008Gu}. The violations decrease as the
external magnetic field gets increased. At the critical point of a quantum
phase transition in this model~\cite{2000Sachdev,2008Gu}, namely $h=1$, the
quantum violation decreases to $1.08866$ and the angles for this violation
are as follows: $\theta^{[1]}_2=\theta^{[3]}_2=-\theta^{[1]}_1=-%
\theta^{[2]}_1 =-\theta^{[3]}_1=\frac{19\pi}{97}$, $\phi^{[1]}_1=%
\phi^{[1]}_2=-\phi^{[2]}_1=\phi^{[2]}_2=\phi^{[3]}_1=\phi^{[3]}_2=\pi/2 $
and $\theta^{[2]}_2=\frac{78\pi}{97}$. The threshold value of $h$ for the
ground state $|\psi_3\rangle_g$ to violate the inequality~(\ref%
{ThreeQubits-BTI}) is $1.375$, indicating that the ground state with $%
h>1.375 $ in this model might not be fully entangled. Consequently,
if one want to detect GME in the model, the external magnetic field
should not surpass this threshold value. It is worthwhile to clarify
that the violations of the inequalities~(\ref{general-QBI}) are only
sufficient but not necessary conditions for GME, namely, some states
that not violate these inequalities may also be genuinely full
entangled. That is the reason why we use the term ``\emph{might not
be fully entangled}" to describe the entanglement of the ground
state with $h>1.375$.

\begin{figure}[tbp]
\includegraphics[width=82mm]{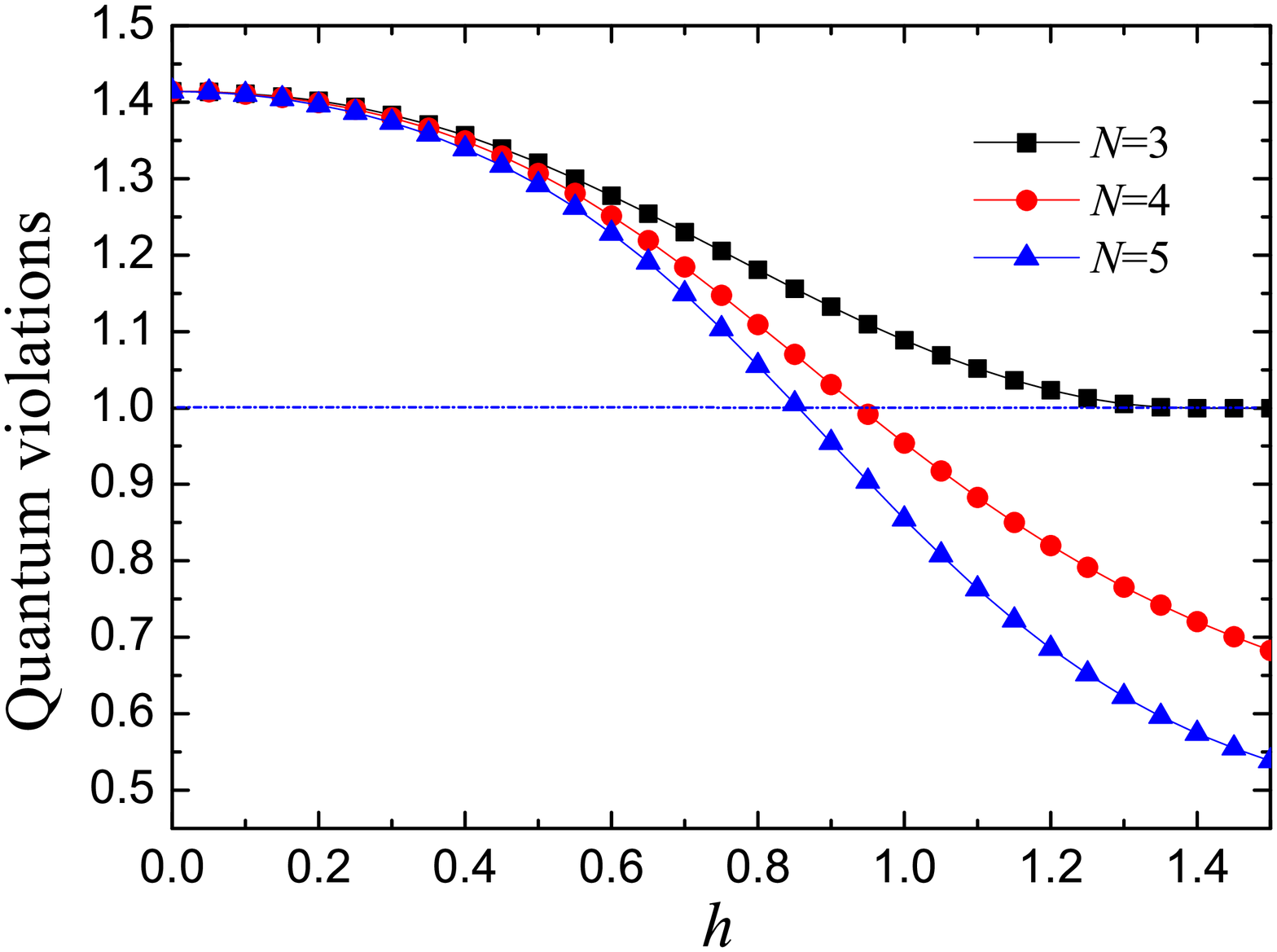}\newline
\caption{(Color online) Quantum violations of the ground states in the
one-dimensional transverse-field Ising model with small number of spins. Here
only the cases $N=3$, $N=4$, and $N=5$ are considered.} \label{QuantumVFig}
\end{figure}

For the four-qubit case, i.e., $N=4$, the ground states of the Hamiltonian $%
H_{4}$ reads:
\begin{eqnarray}
|\psi _{4}\rangle _{g} &=&\frac{1}{\sqrt{\mathcal{N}_{2}}}\left[ \left(
\gamma _{2}-\frac{2\sqrt{2}h(1-\gamma _{3}^{2})}{\gamma _{3}}\right)
|0000\rangle \right.  \\
&+&\left( h+\frac{\gamma _{3}}{\sqrt{2}}\right) |0011\rangle +\frac{4h+2%
\sqrt{2}\gamma _{3}}{2\sqrt{2}\gamma _{3}}|0101\rangle   \notag \\
&+&\left( h+\frac{\gamma _{3}}{\sqrt{2}}\right) |0110\rangle +\left( h+\frac{%
\gamma _{3}}{\sqrt{2}}\right) |1001\rangle   \notag \\
&+&\frac{4h+2\sqrt{2}\gamma _{3}}{2\sqrt{2}\gamma _{3}}|1010\rangle   \notag
\\
&+&\left. \left( h+\frac{\gamma _{3}}{\sqrt{2}}\right) |1100\rangle
+|1111\rangle \right] ,  \notag
\end{eqnarray}%
where $\gamma _{2}=-1+2h^{2}+2\sqrt{1+h^{4}}$, $\gamma _{3}=\sqrt{1+h^{2}+%
\sqrt{1+h^{4}}}$ and $\mathcal{N}_{2}=1+4(h+\frac{\gamma _{3}}{\sqrt{2}}%
)^{2}+\frac{(4h+2\sqrt{2}\gamma
_{3})^{2}}{4\gamma _{3}^{2}}+(\gamma _{2}-\frac{2\sqrt{2}h}{\gamma _{3}}+2%
\sqrt{2}h\gamma _{3})^{2}$ is the normalization constant. A similar analysis
shows that the ground state $|\psi _{4}\rangle _{g}$ violates the
inequalities~(\ref{general-QBI}) with small values of $h$ and the maximal
violation is also $\sqrt{2}$ and occurs at $h=0$. A notable difference
between four-qubit case and three-qubit case is that the violations for
four-qubit ground state decrease faster than that for three-qubit. The
threshold value of $h$ for the ground state $|\psi _{4}\rangle _{g}$ to
violate the inequalities~(\ref{general-QBI}) is $0.935$, which is smaller
than $1.375$, the corresponding threshold values of $h$ for three-qubit
case. The detailed quantum violations as a function of $h$ for the cases $N=3
$, $N=4$, and $N=5$ are illustrated in Fig. \ref{QuantumVFig}. From Fig. \ref%
{QuantumVFig}, we can see that the quantum violations decrease faster and
faster as the system size increase. At the critical point of a quantum phase
transition, i.e., $h=1$, the corresponding ground states for $N=4$ and $N=5$
do not violate the inequalities~(\ref{general-QBI}), which is different from
the $N=3$ case.

For larger systems, it is very difficult to calculate the quantum violations
since there are so many parameters in the maximizing procedure. From the
numerical results above, we believe that GME does exist in the one-dimensional
transverse-field Ising model, but the more particles involved in the model, the
small threshold value of $h$ for the ground states to preserve GME.

It is worthwhile to note that the Bell-type inequalities~(\ref{general-QBI})
originate from the HLNHV models, thus these inequalities are also useful for
the study of genuine multipartite nonlocality (GMNL), which is quite
different from the concept GME (for details, see Ref.~\cite%
{2004Mitchell-three-particle-nonlocality}). Actually, T\'{o}th and
Ac\'{\i}n have presented a concrete example in Ref.
\cite{2006GTrPELocalM-Toth} illustrating that a certain family of
genuine three-qubit entangled states preserve local hidden-variable
models. Any violation of inequalities (\ref{general-QBI}) is a
sufficient condition of GMNL.  Consequently, the quantum violations
of these inequalities in the one-dimensional transverse-field Ising
model also confirm GMNL in this model, implying that the model can
not be described by HLNHV theories.

Recent experiments designed to detect GME are based on the photons and
atom-cavity techniques~\cite{multipartite-entanglement-experiment}. It has been
argued that their claims of experimental confirmation of three- and
four-particle entanglement are questionable~\cite{2001Seevinck}. Fortunately,
the rapid development in the field of Nuclear Magnetic Resonance Quantum
Information Processing (NMR-QIP) has shown that NMR is a valuable and feasible
testing tool for the new ideas in quantum information science (for recent
reviews see Ref.~\cite{2004Vandersypen} and references there in). More than
fifty years of development has put NMR in an unique position to perform complex
experiments and many physical models can be simulated using the NMR
technologies~\cite{2005Peng,2005Negrevergne}. Because of the fundamental
importance of the GME for large scale quantum information processing, further
experimental tests of GME might be widely welcomed. Based on the numerical
results analyzed above, experiments utilizing NMR technologies to confirm GME
in the one-dimensional transverse-field Ising model are nice alternatives.
Actually, the significance of such experiments is at least two-folded: On the
one hand, these experiments might close the loophole problem in the recent
experiments mentioned above and lead to a more comprehensive understanding of
entanglement in the model; On the other hand, this approach can also address
the GMNL problems and might provide experimental evidences concerning the
contradictions between quantum mechanics and HLNHV theories.

In summary, we studied the GME problem in the one-dimensional transverse-field
Ising model based on the Bell-type inequalities. By numerically investigating
the quantum violations of the ground states of this model, we showed evidently
that these ground states are genuinely full entangled when the external
transverse magnetic field is weak. The violations decrease as the external
magnetic field gets stronger and stronger and the more particles involved in
the model, the faster the violations decrease. Our approach also addressed the
GMNL problem. The quantum violations of the Bell-type inequalities
~(\ref{general-QBI}) also confirm GMNL in this model, indicating that the model
can not be described by HLNHV theories. Based on the numerical analysis, we
suggest experiments using the rapid developing NMR technologies to simulate the
one-dimensional transverse-field Ising model be carried out to detect GME and
GMNL in the model.

D. L. Deng is indebted to M. G. Hu for stimulating discussions, critique, and
valuable ideas. S. J. Gu is grateful for the hospitality of Chern Institute of
Mathematics at Nankai University. This work was supported in part by NSF of
China (Grant No. 10605013), Program for New Century Excellent Talents in
University, the Project-sponsored by SRF for ROCS, SEM, and the Earmarked Grant
Research from the Research Grants Council of HKSAR, China (Project No. CUHK
400807).

\end{document}